\documentclass[11pt]{article}
\usepackage{amssymb,latexsym}
\usepackage[dvips]{graphicx} 
\usepackage{url}

\newif\iftemp \tempfalse 
\temptrue

 \catcode`@=11\def\@oddhead{\hbox{}\hfil\rm\thepage}\def\@oddfoot{}
 \def\@evenhead{\hbox{}\hfil\rm\thepage}\def\@evenfoot{}
 \@twosidetrue \catcode`@=12

\iftemp
\DeclareFixedFont{\itshape}{OT1}{cmr}{m}{it}{11}

\fi

\newtheorem{prp}{Proposition}
\newtheorem{lem}[prp]{Lemma}\newtheorem{thm}[prp]{Theorem}
\newenvironment{prf}{\begin{trivlist}\item[\emph{Proof.}]}{\end{trivlist}
  \medskip\par}
\newenvironment{rem}{\begin{trivlist}\item[\emph{Remarks.}]}{\end{trivlist}
  \medskip\par}
\def\prpb{\begin{prp}}\def\prpe{\end{prp}}
\def\lemb{\begin{lem}}\def\leme{\end{lem}}
\def\thmb{\begin{thm}}\def\thme{\end{thm}}
\def\corb{\begin{cor}}\def\core{\end{cor}}
\def\prfb{\begin{prf}}\def\prfe{\end{prf}}
\def\remb{\begin{rem}}\def\reme{\end{rem}}
\def\prpa#1{\label{p:#1}}\def\prpu#1{Proposition~\ref{p:#1}}

\def\thma#1{\label{t:#1}}\def\thmu#1{Theorem~\ref{t:#1}}

\def\seca#1{\label{s:#1}}\def\secu#1{Section~\ref{s:#1}}

\def\itmb{\begin{enumerate}}\def\itme{\end{enumerate}}
\def\itdb{\begin{itemize}}\def\itde{\end{itemize}}
\def\ittb{\begin{description}}\def\itte{\end{description}}
\def\eqnb{\begin{equation}}\def\eqne{\end{equation}}
\def\arrb#1{\begin{array}{#1}}\def\arre{\end{array}}
\def\tabb#1{\par\noindent\begin{tabular}{#1}}
\def\tabe{\end{tabular}\par\noindent}
\def\eqna#1{\label{e:#1}}\def\eqnu#1{(\ref{e:#1})}

\def\QED{\relax\ifmmode\let\@tempa\relax\ifcase\@eqcnt\def\@tempa{& & &}\or
  \def\@tempa{& &}\else\def\@tempa{&}\fi\@tempa $\Box$ \else\hfill $\Box$ \fi}
\def\DDD{\relax\ifmmode\let\@tempa\relax\ifcase\@eqcnt\def\@tempa{& & &}\or
 \def\@tempa{& &}\else\def\@tempa{&}\fi\@tempa $\Diamond$
 \else\hfill $\Diamond$ \fi}

\def\Rom#1{\uppercase\expandafter{\romannumeral#1}}
\def\dsp{\displaystyle}

\def\limf#1{\displaystyle \lim_{#1\to\infty}}

\def\Ccomb#1#2{\setbox0=\hbox{$\displaystyle\mathrm{C}$}\setbox1=\hbox{%
$\scriptstyle #1$}\kern \wd1{\mathrm{C}}_{\kern -1.05\wd0\kern -0.99\wd1{#1}
 \kern 1.15\wd0{#2}}}

\def\clvec#1#2#3{\def\clvecone{#3}\left(\arrb{c} \dsp #1\\ \dsp #2
 \ifx\clvecone\empty\else\\ \dsp #3\fi\arre\right)}

\def\diff#1#2{\dsp\frac{d\,#1}{d#2}}

\def\pderiv#1#2{\dsp\frac{\partial\,#1}{\partial#2}}

\def\le{\leqq} \def\leq{\leqq}\def\ge{\geqq} \def\geq{\geqq}

\def\prb#1{\def\prbone{#1}
  \ifx\prbone\empty{\mathrm{P}}\else{\mathrm{P[\;}}#1{\mathrm{\;]}}\fi}
\def\prbseq#1#2{\def\prbseqone{#2}
  \ifx\prbseqone\empty{\mathrm{P}}_{#1}\ignorespaces
  \else{\mathrm{P}}_{#1}{\mathrm{[\;}}#2{\mathrm{\;]}}\fi}

\def\EEseq#1#2{\def\EEseqone{#2}
  \ifx\EEseqone\empty{\mathrm{E}}_{#1}\else
 {\mathrm{E}}_{#1}{\dsp\mathrm{[\;}}#2{\mathrm{\;]}}\fi}

\def\VVseq#1#2{\def\VVseqone{#2}
  \ifx\VVseqone\empty{\matrm{V}}_{#1}\else
 {\mathrm{V}}_{#1}{\dsp\mathrm{[\;}}#2{\mathrm{\;]}}\fi}

\def\ssN{^{(N)}}

\def\figa#1{\label{f:#1}}\def\figu#1{Fig.~\ref{f:#1}}

\title{
 Mathematical analysis of long tail economy
 using stochastic ranking processes
}
\author{
Kumiko Hattori
\\ 
\small Department of Mathematics and Information Sciences,
\\
\small  Tokyo Metropolitan University, Hachioji, Tokyo 192-0397, Japan.
\\ \small email: \url{khattori@tmu.ac.jp}
\\ \and
Tetsuya Hattori
\\ 
\small Mathematical Institute, Graduate School of Science,
\\
\small  Tohoku University, Sendai 980-8578, Japan.
\\ \small URL: \url{http://www.math.tohoku.ac.jp/~hattori/research.htm}
\\ \small email: \url{hattori@math.tohoku.ac.jp}
} 
\date{\today}

\begin{document}
\maketitle

\begin{center}
ABSTRACT
\end{center}

We present a new method of estimating the distribution
of sales rates of, e.g., book titles at an online bookstore, from 
the time evolution of ranking data found at websites of the store.
The method is based on new mathematical results
on an infinite particle limit of the stochastic ranking process,
and is suitable for quantitative studies
of the long tail structure of online retails.
We give an example of a fit to the actual data obtained from Amazon.co.jp,
which gives the Pareto slope parameter of the distribution of
sales rates of the book titles in the store.

\vspace*{1in}\par
\noindent\textit{Key words:} long tail; online retail; internet bookstore;
 ranking; Pareto
\bigskip\par
\noindent\textit{JEL Classification:} C02
\bigskip\par
\footnotetext{ \noindent\textit{Corresponding author:} 
Tetsuya Hattori, 
\url{hattori@math.tohoku.ac.jp}
\par\noindent
Mathematical Institute, Graduate School of Science, Tohoku University,
 Sendai 980-8578, Japan
\par\noindent
tel+FAX:  011-81-22-795-6391}

\newpage

\section{Introduction.}
\seca{1}

Internet commerce has drastically increased product variety
through low search and transaction costs and nearly unlimited
inventory capacity.  With this new possibility a theory \cite{longtail} has
been advocated which claims that a huge number of poorly selling products
(long tail products) 
that are now available on internet catalogs could make a significant
contribution to the total sales.  In this paper, we refer this 
theory as the possibility of long tail business. 

In studying the possibilities of long tail business,
we need a precise, quick, and costless quantitative method of analyzing
the long tail structure, but there we encounter a problem.
For example, online bookstores have  millions of books
on their electronic catalogues,
but many of the books have average quarterly sales less than $1$.
This means that
if we start collecting the sales record, we will end up,
after waiting for $3$ months, with a list which has ten thousand lines
with $0$ sale and another ten thousand with $1$ sale, and so on.
Moreover, the result will not mean that a particular book with $1$ sale 
has a better potential sales ability than a book with $0$ sale:
A problem characteristic of quantitative analysis of long tail business
is, that for product items of low sales potentials, fluctuations dominate
in the observed data.
Even though we want to suppress fluctuations, 
since each item produces very little profit,
we cannot afford to spend time and money in collecting
extensive data over a long period required from the law of large numbers.

If we hope to estimate the total sales of a store,
we could obtain it from an observation in a short period
with less relative fluctuations, 
thanks to the law of large numbers.
For a revenue officer, this may be sufficient.
But for those who we are interested in the long tail business, 
for example,  an executive running the online store or 
a stockholder waiting for disclosure, as well as  
an observer for research purpose, 
a detailed structure of the contribution of less sold items
would be important.
More specifically, we would like to know the \textit{distribution} of
sales potentials of the products at an online store, such as
the ratios of the number of items with average sales rate
below any given number.
As discussed in the previous paragraph, extracting the
average sales rate of an item would require a long time of observation.
One would then 
consider observing sufficiently many items of relatively low sales
and calculate an average, to suppress statistical fluctuation,
but then one faces a problem of selecting product items of similar
sales potential,
and we come back to the problem of statistical fluctuation
for the data on a single item in the long tail regime.

On web pages, various ranking data can be found.
An example is the sales rankings of books at 
online bookstores such as Amazon.com.  On the web page of each 
book, we see, as well as the title, price, and description of the book,
a number ranging from 1 to several millions which indicates the
book's relative sales ranking at the online store.
In this paper, based on the analysis of 
a mathematical model defined and studied in \cite{HH071,HH072},
we propose a new and simple method, using the ranking data, 
to overcome the problem
of statistical fluctuations of the data on items with low sales
potential.
Our method allows us, by
observing how the sales ranking of a single product develops
with time, 
to reproduce the distribution of sales potentials of all the products
sold at the online store,
free of statistical fluctuations.
Our theory could serve as an efficient and inexpensive method
of a prompt analysis of long tail sales structure.

The plan of the paper is as follows.
In \secu{2} we review the model of stochastic ranking process,
and explain the main theorems in \cite{HH071,HH072}.
To test the applicability of our theory in practical situations,
we apply in \secu{3} the formulas summarized in \secu{2} to
the rankings at Amazon.co.jp.
In \secu{4} we discuss further implications of
the theory of the stochastic ranking process
and possible implications of the results obtained at Amazon.co.jp.

\section{Formulation.}
\seca{2}

In this section, we summarize the main results in \cite{HH071}
on the stochastic ranking process.
It is a simple model that  
describes the time development of sales rankings at online bookstores.

Consider a system of $N$ items (say, book titles), each of which has 
a ranking ranging from $1$ to $N$ so that no two items have the same ranking.
Each item sells at random times.  Every time (a copy of) an item sells, 
the item jumps to rank $1$ immediately.
If its ranking was $m$ before the sale, all the items that
had rank $1$ through $m-1$ just before the sale shift to rank $2$ 
through $m$, respectively.
Thus, the motion of an item's ranking consists of jumps to the top
and monotonous increase in the ranking number between its own sales,
caused by the sales of numerous other items.

We prove that under appropriate assumptions, 
in the limit $N \to \infty$, the random motion of each item's ranking between
sales converges to a deterministic trajectory.  This trajectory can actually
be observed as the time-development of a book's sales ranking at 
Amazon.co.jp's website.
Simple as our model is, its prediction fits well with observation and 
allows the estimation of the Pareto slope parameter. 
We also  prove that the 
(random) empirical distribution of this system (sales rates and scaled
rankings)  converges to a deterministic time dependent distribution.  

To formulate the model mathematically, let us introduce notations and 
state assumptions.
Let $i=1, \cdots N$ be the labels that distinguish the items.
We denote the sales ranking of item $i$ at time $t$ by $X\ssN_i(t)$, 
for  $i=1,2, \cdots , N$.
Assume that a set of initial rankings $x\ssN_{i,0}=X\ssN_i(0)$,
satisfying $x\ssN_i(0) \ne x\ssN_{i'}(0)$ for $i\ne i'$, and 
sales rates $w\ssN_i>0$ are given (non-random).
Namely, items with various sales rates (selling well or poorly)
start with these given initial rankings $x\ssN_{i,0}$, 
and set out to motion according to their sales rates. 
Let $\tau\ssN_{i,0}=0$ and $\tau\ssN_{i,j}$,
$i=1, \cdots , N$, $j=1,2, \cdots $, 
be the $j$-th sales time of item $i$, which is a random variable.
Assume that sales of different items occur independently, and furthermore, 
for each $i$, the time interval between sales 
$\{ \tau\ssN_{i,j+1}-\tau\ssN_{i,j}\}_{j=1,2, \cdots } $
are independent and have an identical exponential distribution to
that of $\tau\ssN_{i,1}$ given by
\[\prb{\tau\ssN_{i,1}\le t} =1-e^{-w\ssN_it},\ \ t\geq 0.\]
A property of exponential distributions implies that 
$w\ssN_i $ corresponds to the average number of sales per unit time.
In the time interval $(\tau\ssN_{i,j},\tau\ssN_{i,j+1})$
the ranking $X\ssN_i(t)$ increases by $1$ every time one of
other items in the tail side of the sales ranking 
(i.e., with larger $X\ssN_{i'}(t)$) sells. 
Thus, the stochastic ranking process is defined as follows:
for $i=1, \cdots , N$,
\itmb
\item
$X\ssN_i(0)=x\ssN_{i,0},$
\item
$X\ssN_i(\tau\ssN_{i,j})=1$, $j=1,2,\cdots$,
\item
for each $i'\ne i$ and $j'=1,2,\cdots$,
if $X\ssN_i(\tau\ssN_{i',j'}-0)<X\ssN_{i'}(\tau\ssN_{i',j'}-0)$ then
$\dsp X\ssN_i(\tau\ssN_{i',j'})=X\ssN_i(\tau\ssN_{i',j'}-0)+1$,
where $\tau\ssN_{i',j'}-0$ means `just before' time $\tau\ssN_{i',j'}$,
\item
otherwise $X\ssN_i(t)$ is constant in $t$.
\DDD
\itme
Since sales rankings are determined by random sales times, 
sales rankings are also random variables.

Let  $\dsp x\ssN_C(t)= \sharp\{ i \mid \tau\ssN_{i,1}\le t \}$,
where $\sharp A $ denotes the number of the elements of a set $A$.
$x\ssN_C(t)$ is the number of the items which has sold at least once by
time $t$.
Note that in the ranking queue of items, the item with rank $x\ssN_C(t)$ 
marks a boundary; 
all the items with  $X\ssN_i(t)\leq x\ssN_C(t)$ (`higher' rankings) 
has experienced a sale, while those with $X\ssN_i(t)> x\ssN_C(t)$ 
(`lower' rankings) have not sold at all by time $t$. 

We can also see   
$x\ssN_C(t)+1$, $0\leq t\leq T$ as the trajectory of the sales ranking 
of an item
that started with rank 1 at time $0$ and has not sold by time $T$.  
It is convenient to consider the scaled trajectory defined 
by $\dsp y\ssN_C(t)=\frac1N x\ssN_C(t)$,
for it is confined in the finite interval $[0,1]$.
The scaled trajectory is random, but 
the following proposition shows that this random trajectory converges to a 
deterministic (non-random) one as $N\to \infty $.

Recall that item $i$ has sales rate $w\ssN_i$.  This determines 
the empirical distribution of sales rate as 
$\dsp\lambda\ssN(dw)=\frac1N \sum_{i=1}^N \delta_{w\ssN_i}(dw)$,
where $\delta_c$ with $c \in {\mathbb R}$ denotes a unit distribution
concentrated at $c$. Namely, for any set $A \subset [0,\infty)$,
\[
\int_A \delta_c (dw)  =\left\{ \arrb{ll}\dsp
1\,, & \mbox{if }\ \ c \in A, \\
\dsp
0\,, & \mbox{if }\ \ c \not\in A.
\arre \right. 
\]
\prpb
\prpa{yC}
Assume that the empirical distribution of sales rate $\lambda\ssN$
converges as $N\to\infty$ weakly to a distribution $\lambda$.
Then 
\eqnb
\eqna{yc}
y\ssN_C(t) \to y_C(t)
\eqne
in probability, where
\eqnb
\eqna{classicstationaryexp}
y_C(t)=1- \int_0^{\infty} e^{-w t} \lambda(dw).
\eqne
\DDD
\prpe
This proposition is a straightforward result of the law of large numbers.
Intuitively, the stochastic process $y\ssN_C$ converges to the
deterministic curve $y_C$ because a trajectory of an item between
the point of its sales
is determined by the independent sales of numerous others
(towards the tail side of the book in observation in the ranking).
The popularity of the observed book is reflected in the length
of sojourn in the sequence before it makes next jump (i.e., ordered
for sales.)
\remb
\itmb
\item
The random variable $y\ssN_C(t)$ converges as $N \to \infty$ 
to a {\it deterministic} 
quantity $y_C(t)$.  It implies that if $N$ is large enough, 
the scaled trajectory provides us with fluctuation-free information.
If we try to know the sales rate of each product by counting the sales
for a certain period of time, we cannot avoid fluctuation.   
The more precise data we want, 
the more time is needed to count the sales, especially for items
that rarely sell, say, once a month.
This proposition ensures that 
by observing the time development of the sales ranking of 
a {\it single} item, we can reproduce the distribution of sales rates, 
free of statistical fluctuation.

\item
$\dsp L(t)= \int_0^{\infty} e^{-wt} \lambda(dw)$  on the right-hand side 
of \eqnu{classicstationaryexp} is the Laplace transform 
of the distribution $\lambda $. 
There is a uniqueness theorem according to which the Laplace transform 
completely determines the distribution \cite{Billingsley}. 
\DDD
\itme
\reme

Intuitively, we can guess that near the top of the ranking, 
there are more items with large sales rates than in the tail regime.
This intuition can be made mathematically precise and rigorous:
\thmb 
\thma{HDL}
Assume the following:
\itmb
\item[(1)] The combined empirical distribution of sales rate and 
the initial scaled sales rankings 
$\dsp y\ssN_{i,0}=\frac1N\, (x\ssN_{i,0}-1)$ 
\[ \mu\ssN_{y,0}(dw\,dy)
=\frac1N \sum_i \delta_{w\ssN_i}(dw)\, \delta_{y\ssN_{i,0}}(dy), \]
converges as  $N\to\infty$ to a distribution 
$\mu_{y,0}(dw)\,dy$
on ${\mathbb R}_+ \times [0,1]$ 
which is absolutely continuous with regard to the Lebesgue measure on $[0,1]$.
\item[(2)] $\lambda(\{0\})=0$D
\item[(3)] $\dsp \int_0^{\infty} w \lambda(dw)<\infty$.
\itme
Then
the combined empirical distribution of sales rate and scaled rankings 
$\dsp Y\ssN_i(t)=\frac1N\, (X\ssN_i(t)-1)$
\[
\mu\ssN_{y,t}(dw\,dy)=
\frac1N \sum_i \delta_{w\ssN_i}(dw)\, \delta_{Y\ssN_i(t)}(dy)
\]
converges as $N\to\infty$ to a distribution 
$\mu_{y,t}(dw)\,dy$ which is absolutely continuous with regard to 
the Lebesgue measure on $[0,1]$.

In particular, the ratio of items with $0<a\leq w \leq b$ and rankings in 
$[0,y]\subset [0,1)$ at time $t$
is given by
\eqnb
\eqna{Tets20070726}
\int_0^y \mu_{z,t}([a,b]) \, dz =\left\{ \arrb{ll}\dsp
\int _a^b (1-e^{-wt_0(y)} )\lambda(dw),
 & y<y_C(t), \\
\dsp
\int_a^b (1-e^{-wt} )\lambda(dw)
+\int _a^b e^{-wt} \int _0^{\hat{y}(y,t)}\mu_{z,0}(dw)\, dz,
 & y>y_C(t),
\arre \right. 
\eqne
where $t_0(y)$ is the inverse function of the strictly increasing
continuous function $y_C(t)$:
\eqnb
\eqna{t0y}
y_C(t_0(y))=y,\ \ 0\le y<1,
\eqne
and $\hat{y}(\cdot,t)$ is the inverse function of 
$\dsp y_C(y,t)=1-\int_y^1 \int_0^{\infty} e^{-wt} \mu_{z,0}(dw)\,dz.$, 
which is a strictly 
increasing continuous function of $y$. 

Furthermore, the trajectory $\frac1N\, X\ssN_i(\tau_{i,j}+t)$, 
time-shifted by $\tau_{i,j}$,
converges as $N\to \infty$ to $y_C(t)$ given in \prpu{yC} 
up to the next jump time ( $0\leq t\leq \tau_{i,j+1}-\tau_{i,j}$ ).
\DDD
\thme
\remb
\itmb
\item
Assumption (1) says that in actual applications we are considering
a long tail economy with a large number of items $N\gg1$,
and that we may regard the empirical distribution $\mu\ssN_{y,0}$ 
at the starting point of observation as a continuous distribution.

\item
Assumption (2) implies  that all the items sell.
With extra notations \thmu{HDL} essentially holds without
Assumption (2), but we will keep it to avoid complications.

This assumption implies that $y_C$ is a  strictly increasing function 
of $t$, and the inverse function $t_0:\ [0,1)\to[0,\infty)$ exists.
Under Assumption (2), 
${y}_C(y,t)$ is a strictly increasing function of $y$, thus the inverse 
$\hat{y}(\cdot,t):\ [y_C(t),1)\to[0,1)$ exists. 

\item
Assumption (3) assures the explicit form of the limit \eqnu{Tets20070726} 
in the following Theorem to hold also for $y=0$.
For $y>0$ the Theorem holds without Assumption (3).
(Hence the only essential assumption is the Assumption (1).)

\item
The last statement in the Theorem implies that by observing
the time development of the ranking $x_C(t)$ of {\it any single item}
from the moment of its sales point ($x_C(0)=0$),
we can, by equating $y_C(t)=x_C(t)/N$ with \eqnu{classicstationaryexp},
obtain the information on the {\it distribution} of sales potential $\{w_i\}$,
of {\it all the items} listed in the rankings.

\item
This Theorem is mathematically nontrivial in the sense that 
a law of large numbers of `dependent' random variable is 
the key to the proof.

It is also known that
$\mu_{y,t}(dw)$ satisfies the following set of partial differential equations:
For any measurable set $A\subset [0,\infty)$,
\[
\pderiv{\mu_{y,t}(A)}{t} +\pderiv{(v(y,t)\, \mu_{y,t}(A))}{y}
=-\int_A w \mu_{y,t}(dw),
\ \ \ 
\pderiv{v}{y}(y,t) =- \int_0^{\infty} w\, \mu_{y,t}(dw).\]

For mathematical details, see \cite{HH071,HH072}.
\DDD
\itme
\reme

In the subsequent sections we
consider the stochastic ranking process as a model
for the rankings found, for example, at the web sites of an online bookstore.
We regard an item in the model as a book title,
and the jump time to rank $1$ as the time that the title is ordered for
sale.
According to the definition of the model,
we assume that each time a book is ordered the ranking of the title
jumps to  $1$, no matter how unpopular the book may be.
At first thought one might guess that such a naive ranking
will not be a good index for the popularity of books.
But thinking more carefully, one notices that well sold books
(items with large $w\ssN_i$, in the model) are dominant
near the head of the ranking, while books near the tail are rarely sold.
Hence, though the ranking of each book is stochastic and has sudden jumps,
the spacial \textit{distribution} of jump rates are more stable,
with the ratio of books with large jump rate high near the top position
and low near the tail position.
Seen from the bookstore's side, it is not a specific book that really
matters, but a totality of book sales that counts, so 
the evolution of \textit{distribution} of jump rate is important.
\thmu{HDL} says that we can make
this intuition rigorous and precise, with an explicit form of the 
distribution when the total number of titles in the catalog of the
bookstore is large (i.e., in the large $N$ limit).

\section{Application to sales analysis of Amazon.co.jp.}
\seca{3}

In this section, we give an explicit example of how the theoretical framework
in \secu{2} could be applied to realistic situations.
We will focus on the sales ranking data found at the websites of
Amazon.co.jp, the Japanese counterpart of the online bookstore Amazon.com.

We first give in \secu{30} 
a brief explanation about the sales ranking number
found at the web pages for Japanese books at Amazon.co.jp,
and summarize in \secu{31} the method of applying \secu{2} to 
actual ranking data, and give an explicit result of statistical fits of the
distribution of sales rate of the books at the online bookstore.

\subsection{Amazon.co.jp book sales ranking.}
\seca{30}

The web sites of Amazon (irrespective of countries) have 
a web page for each book title, where we find, as well as its 
title, author and price, 
a number which represents the sales ranking of the book.
It has been noticed \cite{CG,BSH} that
this number serves as an important data for
quantitative studies of the economic impact of online bookstores.
This is because the number reflects the sales rate of the book,
and especially in the situation that, in terms of \cite{BSH},
'Internet retailers are extremely hesitant about releasing
specific sales data', it can be one of the scant data publicly available.

We refer to \cite{CG} for general structure of the web pages,
and to \cite{Rosenthal} for a summary based on apparently a long and
extensive observation of the ranking number at Amazon.com, and in particular, 
discussion on its relation to the actual sales of the book at Amazon.com.
Here we focus on observed facts about the time evolution of ranking numbers
at Amazon.co.jp.
Firstly, it is said that Amazon.com adopts an involved definition
of the ranking numbers than the stochastic ranking process.
Secondly, Amazon.co.jp is easier for the authors to find appropriate
data (it is our home country).

If we keep observing the ranking number of a book,
we soon notice that it is updated once per hour regularly.
For a relatively unpopular book title, 
the corresponding ranking number increases steadily and smoothly
for much of the time as the number is updated, 
but once in a while we see a sudden
jump to a smaller number around ten thousand.
This happens when a copy of the book is ordered for purchase,
which can be checked by personally ordering a copy at Amazon website;
at the update time which is $1$ -- $2$ hours after the order, 
the ranking number is observed to jump.
Actually, except for the top ten thousand sellers out of a few million
Japanese book titles catalogued at Amazon.co.jp, 
a book sells less than $1$ per hour on average,
hence the qualitative motion just described hold for $99$ percent of the book
titles at Amazon.co.jp.

Note that this behavior of the time evolution of a ranking number is similar
to that of stochastic ranking model in \secu{2}.
The correspondence is also natural from an observation by \cite{Rosenthal}
that the Amazon's ranking number system `is based almost entirely on 
{``what have you done for me lately''}'.
For seldom sold books,
any natural definition of the ranking number satisfying such a criterion
would be in the order of latest sales time, because
any sales record before the latest one should be further remote past 
and would have only a small effect on any reasonable definition of 
the ranking number.
Hence the definition of the stochastic ranking process in \secu{2},
even though it may have sounded over-simplified,
has a chance of being a good theoretical basis for modelling
the ranking numbers on the web, especially for 
probing a large collection of titles in the long tail regime of the catalog, 
which is of interest in this paper.

If we further assume as usual that the point of sales are random,
then we will have a full correspondence between the stochastic ranking model
and the time evolutions of ranking numbers at Amazon.co.jp.
Based on the correspondence, we give,
in the next subsection \secu{31}, explicit formulas
which relate a time evolution of a ranking number $x_C(t)$ to
a distribution of average sales rate of the book titles at the bookstore,
and then using the formulas we give results of fits with observed data.

\subsection{Stochastic ranking process analysis of book sales ranking.}
\seca{31}

We start with a standard assumption, as, for example, in \cite{CG,BSH},
that the probability distribution of book sales rate is a Pareto distribution
(also called a power law or a log--linear distribution).
In the notations of \secu{2} this means that we assume the probability
measure $\lambda$ to be 
\eqnb
\eqna{Paretolambda}
\lambda([w,\infty))= \left\{ \arrb{ll} \dsp \left(\frac{a}{w}\right)^b,
 & w\> a, \\ 1, & w<a, \arre \right.
\eqne
where $a$ and $b$ are positive constants.
Its probability density function is given by
\eqnb
\eqna{Paretolambdadensity}
\diff{\lambda}{w}(w) = \left\{ \arrb{ll} \dsp\frac{ba^b}{w^{b+1}}, & w\ge a,
\\ 0, & w<a. \arre \right.
\eqne
In terms of books, $w$ denotes the average sales rate of a book on the list
of a bookstore;
a book with $w$ sells on average in the long run $w$ copies per unit time.
$\lambda$ is the distribution of $w$; for example,
$\lambda([w,\infty))$ is the ratio of the number of book titles with sales 
rate $w$ or more to the total number of titles.
Alternatively we could start with another (discrete) formulation of the Pareto
distribution
\eqnb
\eqna{Paretodiscrete}
w_i=a\left(\frac{N}{i}\right)^{1/b},\ \ i=1,2,3,\cdots,N,
\eqne
where the constant $a$ in \eqnu{Paretodiscrete}
(or in \eqnu{Paretolambda})
denotes the lowest positive sales rate among the book titles
at the store. 
Note that the books that never sell should be omitted in
applying our theory.
$N$ is the total number of such titles as actually sell
catalogued at the online bookstore,
and $w_i$ is the average sales rate of the $i$-th best seller.
The ratio of titles with $w$ or more average sales rate is then
\[ \frac1N \sharp\{ i \mid w_i\ge w \}
 = \frac1N \sharp\{ i \mid i\le N \,\left(\frac{a}{w}\right)^b \}
 = \left(\frac{a}{w}\right)^b, \]
for $w\ge a$, reproducing \eqnu{Paretolambda}.

The exponent $b$ ($-1/b$ corresponds to the Pareto slope parameter)
is crucial in the analysis of economic impact
of the retail business in question. In fact, previous studies using
the ranking numbers at the online bookstores \cite{CG,BSH}
use the data for extracting the exponent $b$,
which then was used 
to study various aspects of economic impact of the online bookstores.
An intuitive meaning of the exponent $b$ can be seen, for example,
by taking ratio of \eqnu{Paretodiscrete} for $i=1$ and $N$, to find
\eqnb
\eqna{Paretobimplication}
\frac{w_1}{w_N}=N^{1/b},
\eqne
which roughly says that for large $N$ if $b$ is small
then $w_1$ is very large compared to $w_N$, so that
the greatest hits dominate the sales,
while if $b$ is large the contributions are more equal,
and since there are many unpopular titles,
their total contribution to the sales may dominate
(the `long tail' possibility).
We will discuss further on the implications of 
the parameter $b$ in \secu{4}.

Our method of obtaining the parameters $a$ and $b$ 
is to observe a time development of the ranking of any single book title, 
which contains information of $\lambda$, with statistical fluctuations
strongly suppressed.
(One may be curious why a data from a single title could have 
fluctuation suppressed.
This is because the time development of the ranking,
during the book in question is not sold,
is a result of the total sales of the the large amount of titles 
in the tail side of the observed book in the catalog of an online bookstore,
hence the statistical fluctuation is suppressed by a law-of-large-numbers
mechanism.
This is a practical meaning of the deterministic motion appearing as
an infinite particle limit stated in \secu{2}.)
Substituting \eqnu{Paretolambda} in \eqnu{classicstationaryexp} we have
\eqnb
\eqna{classicstationaryexpPareto0}
y_C(t)=1-ba^b \int_a^{\infty} e^{-w t} w^{-b-1} dw
=1-b(at)^b \Gamma(-b,at),
\eqne
where $\Gamma$ is the incomplete Gamma function defined by
$\dsp \Gamma(z,p)=\int_{p}^{\infty} e^{-x} x^{z-1} dx. $
Since $b$ is positive $\Gamma(-b,at)\to \infty$ as $t\to0$.
This divergence is mathematically harmless because of the factor $t^b$, but
from a practical point of view, it is convenient to
use the integration-by-parts formula
\eqnb
\eqna{Gammarecursion}
\Gamma(z,p)=-z^{-1}p^{z}e^{-p} +z^{-1} \Gamma(z+1,p)
\eqne
to obtain
\eqnb
\eqna{classicstationaryexpPareto}
y_C(t)=1-e^{-at}+ (at)^b \Gamma(1-b,at).
\eqne
This formula is satisfactory for $0<b<1$.
For $1<b<2$ use \eqnu{Gammarecursion} again to obtain
\eqnb
\eqna{classicstationaryexpPareto2}
y_C(t)=
1-(1-\frac{at}{b-1})\, e^{-at} - \frac{(at)^b}{b-1}\, \Gamma(2-b,at).
\eqne
In principle, we may perform integration by parts as many times as required,
though we did not come across values $b\ge 2$ in the literature or 
in our data.
For $b=1$, we need a slightly different formula with 
`logarithmic corrections', 
but we have not observed any practical evidence that 
the exact value of $b=1$ occurs, so we will always assume $b\ne 1$
in the following, to simplify the formulas.

Note in particular, that \eqnu{classicstationaryexpPareto} implies that
for $b<1$ we have a concave time dependence for short time,
\[
y_C(t)=(at)^b\Gamma(1-b,0)+o(t^b),
\]
while \eqnu{classicstationaryexpPareto2} implies that
for $b>1$ we have linear short time dependences.
According to the results in \secu{2},
$y_C(t)$ is the relative position (i.e., $0\le y_C(t)<1$) at time $t$ 
in the ranking of the title
which was at the top position (i.e. sold) at $t=0$.
The corresponding ranking number $x_C(t)$ is given by
\eqnb
\eqna{xclassicstationaryexpPareto}
x_C(t)\simeq N\, y_C(t)=N\,(1-e^{-at}+ (at)^b \Gamma(1-b,at)).
\eqne
where $N$ is the total number of the catalogued titles that actually sell.
We cannot control subleading order in $N$
because of the statistical fluctuations.
(The limit theorems in \secu{2} assures that the leading order
is free of statistical fluctuations.)
However, since Amazon has a huge `electronic bookshelf' of order $N=O(10^6)$,
we will omit the statistical fluctuations of relative order
 $O(\sqrt{N}^{-1})=O(10^{-3})$.

Incidentally, we can alternatively start from \eqnu{Paretodiscrete} and use
the empirical distribution
$\dsp \frac1N \sum_{i=1}^N \delta_{w_i}$ for $\lambda$,
where $\delta_w$ is a unit distribution concentrated at $w$.
Then from \eqnu{classicstationaryexp} we have, by elementary calculus,
\[
y_C(t)=1- \frac1N \sum_{i=1}^N e^{-a (N/i)^{1/b} t} 
= 1- \int_a^{\infty} e^{-wt} 
ba^b \int_a^{\infty} e^{-w t} w^{-b-1} dw + O(N^{-1}),
\]
reproducing \eqnu{classicstationaryexpPareto0}.

Before closing this subsection, we recall that
\eqnu{classicstationaryexp} implies that
the ranking of an item is, as a function of time $t$, 
essentially the Laplace transform
of the underlying distribution $\lambda$ of the jump (sales) rates.
If we have a accurate and long enough ranking data 
(i.e., observation of the time evolution of the ranking $x_C(t)$
for a very long period and with very fine intervals),
the uniqueness of inverse Laplace transform assures in principle
the determination of $\lambda$ non-parametrically, i.e., without assumptions
on $\lambda$ such as assuming Pareto distribution \eqnu{Paretolambda}.
This approach however requires a very fine data,
because the Laplace transform has smoothing effect through $e^{-wt}$
factor, and a small irregular differences in the Laplace transform
could result in a large difference in the original function.
In the case of Amazon.co.jp, which we see in \secu{32},
the ranking is updated only once per hour
and we cannot expect fine enough data (as is also the case of Amazon.com),
so we will follow a standard approach assuming
a Pareto distribution for $\lambda$.
(Needless to say, the managers in the Amazon company have access to 
precise real-time data, hence
our methods will help them analyze and plan
the inventory controls and evaluate the sales.)

If long tail economy expands in the future, and our methods turn out
to be of practical use, it would be preferable to have
real time spontaneous updates of the ranking data,
which will make our methods more efficient and accurate.
(It will not cost any more than the current Amazon's ranking data
updates with hourly intervals; in fact, 
the title listings at the 2ch.net adopt such algorithms \cite{HH072}.)

\subsection{Results from Amazon.co.jp.}
\seca{32}

By performing a statistical fit to \eqnu{xclassicstationaryexpPareto} 
of ranking time evolution data, 
we can in principle obtain the parameters $a$ and $b$ which determine the
distribution of average sales rates of the book titles at Amazon.co.jp.
In the practical situations, it turns out that 
the total number $N$ of the book titles
also needs to be determined from the data.

We are aware that Amazon.co.jp publicizes at their website 
the total number of book titles on their catalog, which can be 
reached  
by making an unconditioned search at the Amazon website.
However, 
the book catalogs at Amazon websites contain books
which are not available and therefore do not sell,
hence, as we noted below equation \eqnu{Paretodiscrete}
while describing the Pareto distribution,
should be discarded from our analysis.
We have experienced more than once that we order a book at the website
and receive a note after a while that the book has not been found
and that the order is cancelled. At the same time, we observe
the ranking number of that cancelled title making jumps to the tail side.
We thus realize that the claimed number of titles
at the website contains those with $w=0$ and is therefore
strictly larger than what we should use for $N$ in our formulation.
As an explicit example,
the number from Amazon.co.jp search results was 2,587,571
on Oct.~4, 2007, while our fits indicates 
$N$ to be strictly less than $1$ million (see \eqnu{low2}).
\begin{figure}[hbt]
\begin{center}
\includegraphics[scale=1.0]{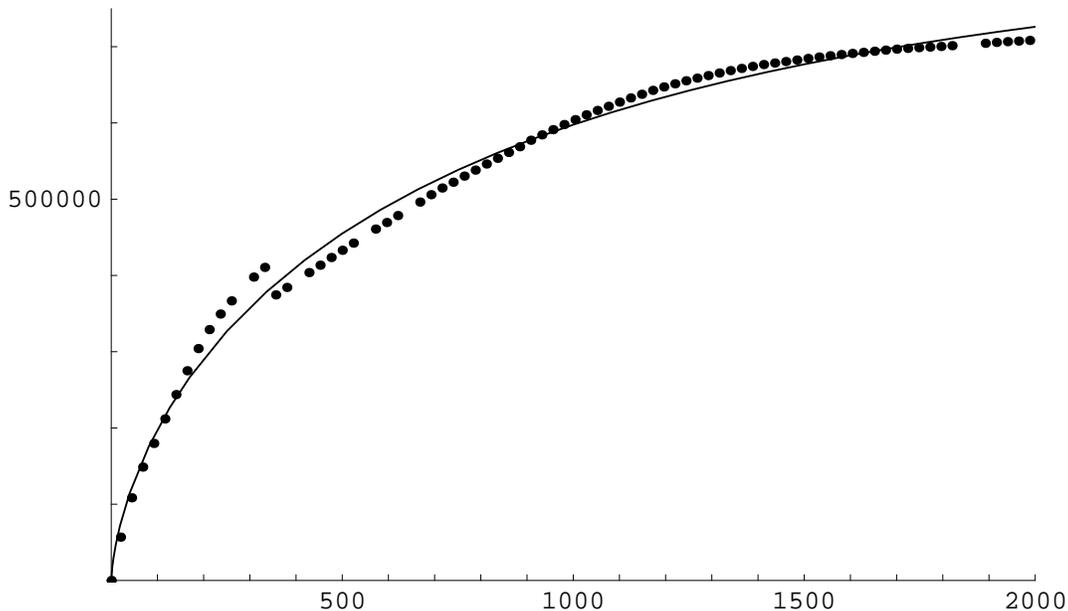}
\caption{A long time sequence of data from Amazon.co.jp.
The solid curve is a theoretical fit.
Horizontal and vertical axes are the hours and ranking, respectively.}
\figa{low2}
\end{center}
\end{figure}
Now we turn to our results of observation.
The plotted $n_d=77$ points in \figu{low2} show 
the time evolution of the ranking of a book we observed
between the end of May, 2007 (at which point the book was 
ordered for sales) and mid August, 2007 (at which point the book
was bought again).
The solid curve is a least square
fit of these points to \eqnu{xclassicstationaryexpPareto}.
The best mean-square fit for the parameter set $(N,a,b)$ is:
\eqnb
\eqna{low2}
(N^*,\ a^*,\ b^*)=
(8.57\times 10^5,\ 3.939\times 10^{-4},\ 0.6312).
\eqne

Note that $N^*$ is large, hence the fluctuations arising from
randomness in the sales are relatively suppressed 
($O(1/\sqrt{N^*})=O(10^{-3})$), as expected,
while the number is smaller than that found by performing a search 
at the Amazon website
($8.57\times 10^{5} < 2.6\times 10^6$),
so that a fit of $N$ is necessary.
$a^*$ is in units of $[1/hour]$ and corresponds to
$3.5$ months for $1/a^*$, which is longer than 
the interval of observation ($2.5$ months).
Our method allows the determination of time constants
longer than the interval of observation because
there are a large amount of (mostly unpopular) titles
which theoretically allow a law-of-large-numbers mechanism.
(The obtained value of $a^*$ does not mean that there are no books at all
which sells, say, only one copy a year on average;
it says that such books are much less than would be expected from
a log-linear (Pareto) distribution and 
have a negligible economic impact.)

The total variance $\chi^2$ of the data from this fit is
$\chi^2=1.599\times 10^{10}$, hence the statistical
fluctuation $\Delta y_C$ of the relative ranking is 
roughly of order
\[ \Delta y_C = \frac1N\Delta x_C\sim \frac1N\sqrt{\chi^2/n_d}=0.02. \]
This seems a little larger than an expectation from the Gaussian fluctuation
which would be of order $1/\sqrt{N}=10^{-3}$.
\figu{low2} suggests that a possible reason of the deviations of data from
the fit is caused by a small jump at about $t=300$ hours.
We suspect this as a result of inventory controls at the web bookstore,
such as unregistering books out of print.
Apparently, Amazon.co.jp in the year 2007 was updating their
catalogs manually and only occasionally,
making it a kind of unknown time dependent external source for our analysis.

\begin{figure}[hbt]
\begin{center}
\includegraphics[scale=1.0]{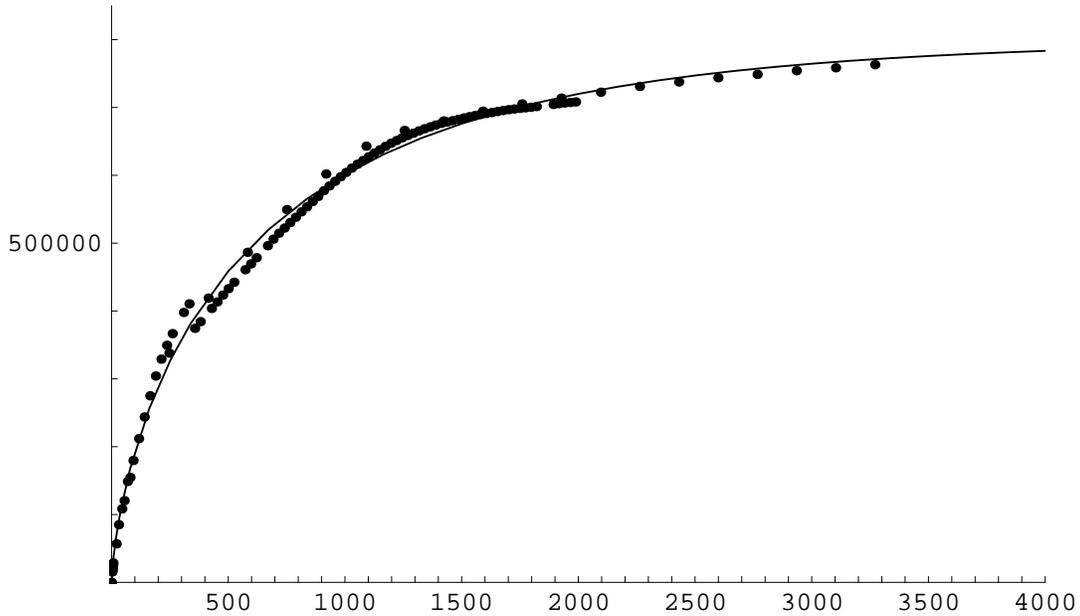}
\caption{Two long time sequence of data from Amazon.co.jp.
One sequence with $77$ points is the data in \protect\figu{low2},
another one with $27$ points.
The solid curve is a theoretical fit to the $77+27$ data.
Horizontal and vertical axes are the hours and ranking, respectively.}
\figa{low24}
\end{center}
\end{figure}
Concerning the stability of the parameters, 
we made another series of observation between
November, 2007 and March, 2008.
This time, having less time to spare we recorded only once a week
resulting in $n_d=27$ points.
The solid curve in \figu{low24} is a least square
fit of the combined $27$ points and the $77$ points in \figu{low2}
to \eqnu{xclassicstationaryexpPareto}.
The best mean-square fit for the parameter set $(N,a,b)$ is:
\eqnb
\eqna{low24}
(N^*,\ a^*,\ b^*)=
(8.00\times 10^5,\ 5.803\times 10^{-4},\ 0.7959).
\eqne
$\chi^2=2.0111\times 10^{10}$ ($\Delta y_C \sim 0.02$)
effectively remained same as \eqnu{low2}.
The parameters have changed somewhat;
change in the total number of active books $N^*$ is not large
(about 7\%), $1/a^*=2.4$ months which is somewhat shorter than \eqnu{low2}.
The exponent $b^*$ is larger, but note that 
we again have exponent $b$ strictly less than $1$.

Though we clearly and consistently have $b<1$ 
(also seen from the concave figure in \figu{low2} and \figu{low24}), 
its value has changed.
The change of $N^*$ and $a^*$
between \eqnu{low2} and \eqnu{low24} is consistent
with a hypothesis that Amazon.co.jp performed inventory controls
(as they should do) and got rid of books with low sales
between the two series of observations,
so one explanation is that the exponent $b$ also changed.
Another possible reason is that the new data of once per week
are too sparse and that we need finer data for stable fits.
In fact, as pointed out at the end of \secu{31},
a fit to the distribution may be sensitive to small changes
in the ranking data, and 
a data finer than once per week may be required.
This problem could be overcome by automated data acquisition
through computer programming.

The values $b^*=0.6312$ in \eqnu{low2} and
$b^*=0.7959$ in \eqnu{low24} are both less than $1$. 
The result, $b^*<1$ obtained from our data may also be 
convincing by a look at \figu{low2} and \figu{low24}, because, 
as we noted below \eqnu{classicstationaryexpPareto2},
the short time behavior of the ranking is proportional to $t^b$
for $b<1$ (which implies the graph is tangential to the ranking axis),
while is linear for $b>1$.
Previous studies \cite{CG,BSH} adopt values $b>1$.
(The correspondences of the notations are
 $b=-1/\beta_2$ for \cite{BSH} and $b=\theta$ for \cite{CG}.
In statistics textbooks $b=\alpha$ and $a = 1/\beta$ are also used.)
According to what we remarked below \eqnu{Paretobimplication},
this implies that, in general, the economic impact of
keeping unpopular titles at online bookstores
may be overestimated in the previous studies.
We will continue on this point in \secu{4}.

\section{Discussions.}
\seca{4}

\subsection{Formulas for the long tail structure of online retails.}
\seca{41}

In \secu{3} we dealt with an application of a formula
\eqnu{classicstationaryexp} in a practical situation,
a prediction on the time evolution of the ranking of a book.
The theoretical framework in \secu{2}, introducing the main results 
of \cite{HH071}, contains more than this, and predicts
the total amount of sales (per unit time) that 
could be expected from the items (e.g., books, in the case of an
online bookstore) on the tail side of any given ranking number $m\le N$.

Note that this is not equal to the total contribution to the sales
from the tail side aligned in order of potential (average) sales rate, which is
$\dsp \sum_{i=m}^N w_i$ in the notations in \secu{3}.
This is because, since the ranking number jumps to the head 
each time the item sells at a random time, and since there are
a very large number of items ($N\gg 1$),
we always have some lucky items with low potential sales around the
head side of the rankings,
and according to a similar argument, 
we also must have some `hit' items towards
the tail side.
The main theorem in \cite{HH071}, as explained in \secu{2},
states that the ratio of such (un-)lucky items
having ranking numbers very different from those expected from their
potential sales ability $w_i$ is non-negligible even in the $N\to\infty$
limit.

An explicit formula can be derived from \eqnu{Tets20070726}.
Note that \eqnu{classicstationaryexp} and Assumption (2) for \thmu{HDL}
imply $\limf{t} y_C(t)=1$, hence after a sufficiently long time since the
start of the bookstore and its ranking system, one may assume that the ranking
reaches a stationary phase and the first
equation in \eqnu{Tets20070726} holds for all $0\le y<1$.
Letting $a=w$ and $b=w+dw$ in \eqnu{Tets20070726} we have
\eqnb
\eqna{salesrankingr1r2pre}
\int_{z\in[0,y]} \mu_{z,t}(dw) \, dz = (1-e^{-wt_0(y)})\,\lambda(dw).
\eqne
Let $0<r_1<r_2\le 1$, and denote by $\tilde{S}(r_1,r_2)$ the contribution
to the total
average sales per unit time from the items with ranking number
between $r_1N$ and $r_2N$. 
For a very large $N$, we may let $N\to\infty$ and use \eqnu{salesrankingr1r2pre}
to find
\eqnb
\eqna{salesrankingr1r2}
\arrb{l}\dsp
\limf{N} \frac1N \tilde{S}(r_1,r_2)
 = \int_{(w,z)\in[0,\infty)\times[r_1,r_2]} w\mu_{z,t}(dw)\,dz
\\ \dsp
 = \int_{(w,z)\in[0,\infty)\times[0,r_2]} w\mu_{z,t}(dw)\,dz
 - \int_{(w,z)\in[0,\infty)\times[0,r_1]} w\mu_{z,t}(dw)\,dz
\\ \dsp
=\int_0^{\infty} w(e^{-w t_0(r_1)} -e^{-w t_0(r_2)}) \,\lambda(dw).
\arre
\eqne
This is valid for an arbitrary sales rate distribution $\lambda$;
for the Pareto distribution \eqnu{Paretolambdadensity} we have,
using the incomplete Gamma function as in \eqnu{classicstationaryexpPareto0},
\eqnb
\eqna{salesrankingr1r2Pareto}
\limf{N} \frac1N \tilde{S}(r_1,r_2)
=
ab\,( \Gamma(1-b,q(r_1))\,q(r_1)^{b-1}
- \Gamma(1-b,q(r_2))\,q(r_2)^{b-1} ),
\eqne
where $q(r)=a\, t_0(r)$ is given by
\eqnu{t0y} with \eqnu{classicstationaryexpPareto}:
\eqnb
\eqna{t0yPareto}
r=1-e^{-q(r)}+ q(r)^b\, \Gamma(1-b,q(r)).
\eqne
For $1<b<2$, a better expression using \eqnu{Gammarecursion}
as in \eqnu{classicstationaryexpPareto2} would be
\eqnb
\eqna{salesrankingr1r2Paretobgt1}
\limf{N} \frac1N \tilde{S}(r_1,r_2)
=
\frac{ab}{b-1}\,(e^{-q(r_1)}- \Gamma(2-b,q(r_1))\,q(r_1)^{b-1}
-e^{-q(r_2)}
+ \Gamma(2-b,q(r_2))\,q(r_2)^{b-1} ),
\eqne
with
\eqnb
\eqna{t0yParetobgt1}
r=1-e^{-q(r)}\,(1-\frac{q(r)}{b-1})- \frac{q(r)^b}{b-1}\, \Gamma(2-b,q(r)).
\eqne

$\tilde{S}(r_1,r_2)$ is to be compared with the contribution $S(r_1,r_2)$ 
to the total average sales per unit time from the items $i$ 
between $r_1N$ and $r_2N$
ordered in decreasing order of potential sales rate $w_i$,
as in \eqnu{Paretodiscrete}. We have,
\eqnb
\eqna{salesr1r2}
\arrb{l}\dsp
\limf{N} \frac1N S(r_1,r_2)=\limf{N} \frac1N \sum_{i=r_1N}^{r_2N} w_i
=\limf{N} \frac1N \sum_{i=r_1N}^{r_2N} a\left(\frac{N}{i}\right)^{1/b}
= a \int_{r_1}^{r_2} x^{-1/b}dx
\\ \dsp
 = \frac{ab}{b-1} (r_2^{(b-1)/b}-r_1^{(b-1)/b}).
\arre
\eqne

Note that $q(0)=0$ and $q(1)=\infty$.
The latter is from \eqnu{classicstationaryexpPareto0}:
\[
r=1-b q(r)^b\, \Gamma(-b,q(r))=1-b\int_1^{\infty}e^{-q(r)y}y^{-b-1}dy.
\]
The last term is a convergent integral for $b>0$, which 
is proved by \eqnu{t0yPareto} for $0<b<1$ and
by \eqnu{t0yParetobgt1} for $1<b<2$.
It converges to $0$ as $q(r)\to\infty$.

The special case of $r_2=1$ corresponds to the contribution from the 
tail side in the ranking for $\tilde{S}(r,1)$ and the 
tail side in the potential sales rate for $S(r,1)$ (the `long tail'),
which are (after some elementary calculus as above)
\eqnb
\eqna{salesrankingr1Pareto}
\arrb{l}\dsp
\limf{N} \frac1N \tilde{S}(r,1)
=ab\, \Gamma(1-b,q(r))\,q(r)^{b-1}
\\ \dsp
=\frac{ab}{b-1}\,(e^{-q(r)}- \Gamma(2-b,q(r))\,q(r)^{b-1}),
\arre
\eqne
with $q(r)$ given by \eqnu{t0yPareto} or \eqnu{t0yParetobgt1},
and
\eqnb
\eqna{salesr1}
\limf{N} \frac1N S(r,1) = \frac{ab}{b-1} (1-r^{(b-1)/b}).
\eqne

Concerning the contributions from the head side (`great hits'),
we note that the cases $b>1$ and $b<1$ are different.
This is easy to see in \eqnu{salesr1r2}, where we find
$\dsp \lim_{r_1\to +0} \limf{N} \frac1N S(r_1,r_2)=\infty$
if $b<1$, while for $b>1$, we can safely take $r_1\to 0$ limit to find
\[
\limf{N} \frac1N S(0,r) = \frac{ab}{b-1} r^{(b-1)/b}.
\]
This quantity
represents an average sales rate per unit time per unit item, 
which is finite for the realistic situations.
For $b<1$ great hits dominate in the total sales,
which theoretically becomes infinitely large as $N\to\infty$
(see \eqnu{Paretodiscrete}), while
for $b>1$ all the items contribute non-trivially, and that with
a large number of items, the contribution from the `long tail' would
dominate, which intuitively explains the difference in the behavior.
The divergence is a result of $N\to\infty$ limit.
We will consider cases $b>1$ and $b<1$ separately
and discuss the implication of the value of $b$ in detail.

\subsection{Implications of the Pareto exponent $b$.}
\seca{42}

We noted at the end of \secu{41} and also below \eqnu{Paretobimplication}
that large $b$ means that the `long tail' is important
while small $b$ means that great hits dominate.
Intuitively,
there are $O(1)$ great hits and $O(N)$ long tail items,
so the ratio of the contribution of the former to the latter
is, using \eqnu{Paretobimplication}, 
$\dsp O(\frac{w_1\times 1}{w_N\times N})=N^{1/b-1}$,
hence when the total number of items $N$ is large,
the dominant contribution to the total sales change
between $b>1$ and $b<1$.

\subsubsection{Case $b>1$: The long tail economy.}
\seca{421}

Let $b>1$ and assume $N$ is large.

For $0\le r\le 1$, 
the contribution to the total sales per unit time of the
$N(1-r)$ items (out of the total $N$) with \textit{low sales potentials}
is given by \eqnu{salesr1}:
\eqnb
\eqna{totalsalestailinpotentialbgt1}
S(r,1) \simeq \frac{Nab}{b-1} (1-r^{(b-1)/b}).
\eqne
In particular, the total sales per unit time at the online store is
\eqnb
\eqna{totalsalesbgt1}
S_{tot}=S(0,1) \simeq \frac{Nab}{b-1}\,.
\eqne
Subtraction gives us the total sales amount 
from the $Nr$ top hits per unit time:
\eqnb
\eqna{totalsalesheadinpotentialbgt1}
S(0,r) \simeq \frac{Nab}{b-1} r^{(b-1)/b}.
\eqne

Similarly, \eqnu{salesrankingr1Pareto} gives the
contribution to the total sales per unit time from the
$N(1-r)$ items in the \textit{tail side of the ranking}:
\eqnb
\eqna{totalsalestailinrankingbgt1}
\arrb{l}\dsp
\tilde{S}(r,1)
\simeq Nab\, \Gamma(1-b,q(r))\,q(r)^{b-1}
=\frac{Nab}{b-1}\,(e^{-q(r)}- \Gamma(2-b,q(r))\,q(r)^{b-1});
\\ \dsp
r=1-e^{-q(r)}\,(1-\frac{q(r)}{b-1})- \frac{q(r)^b}{b-1}\, \Gamma(2-b,q(r)).
\arre
\eqne
In particular, noting $q(0)=0$ and
\[
\Gamma(1-b,q)\,q^{b-1}=\int_1^{\infty} e^{-qy} y^{-b}\,dy
 \to \int_1^{\infty} y^{-b}\,dy=\frac1{b-1}\,,\ q\to 0,
\]
we have $\dsp \tilde{S}(0,1)=\frac{Nab}{b-1}$ for $b>1$, which is equal to
\eqnu{totalsalesbgt1} as expected, because all the items in the store
are listed on the ranking.
Subtraction gives us the total sales amount from the top $Nr$ items in the
ranking (at any given time, if the ranking is stationary) per unit time:
\eqnb
\eqna{totalsalesheadinrankingbgt1}
\tilde{S}(0,r)
\simeq Nab\, (1-\Gamma(1-b,q(r))\,q(r)^{b-1})
=\frac{Nab}{b-1}\,(1-e^{-q(r)}+ \Gamma(2-b,q(r))\,q(r)^{b-1}).
\eqne

The large $b$ implies that there is a good chance in the long tail business.
For example, for a extreme case of $b=2$,
\eqnu{totalsalesheadinpotentialbgt1} implies
$S(0,0.2)/S(0,1) \simeq \sqrt{0.2}\simeq 0.447$,
so that top 20\% of hit items contribute only 45\% of total sales,
far less than 80\% , 
challenging the widespread `20--80 law'.
This is, however, too extreme, and we should use realistic values.
Concerning the analysis based on the rankings of Amazon.com,
Chevalier and Goolsbee \cite{CG} 
explored a number of sources of information,
including their own experiment, and
obtained values for the exponent $b$ ranging from $0.9$ to $1.3$,
and adopted the value $b=1.2$ for their subsequent calculations,
to find, for example, that the online bookstores have more price
elasticity  than the brick-and-mortar bookstores and have a significant
effect on the consumer price index.
Brynjolfsson, Hu, and Smith \cite{BSH} 
uses $b=1.15$ ($-1/b=\beta_2=-0.871$ in their notations),
to evaluate the increase in consumer welfare by the introduction
of large catalogues of books by the online bookstores.
They also quote the values in \cite{CG} and report a result of similar
experiment to obtain $b=1.09$.
For $b=1.2$ and $b=1.15$ we have
$S(0.2,1)/S_{tot}\simeq 0.235$ and 
$S(0.2,1)/S_{tot}\simeq 0.189$, respectively,
behaving more or less like `20--80 law'.
Of course, we are considering $N$ of order of million (or more,
with the advance in the web 2.0 technologies and online retails 
expected in the close future)
distinct items as in \eqnu{low2} or \eqnu{low24}, 
and top 20\% also means a large number.
The term `possibility of the long tail business' makes sense for
$b>1$, in the sense that, with 
a drastic decrease in the cost for handling a large inventory
through online technology, a retail with a 
million items on a single list produces a large profit.

\begin{figure}[hbt]
\begin{center}
\includegraphics[scale=1.0]{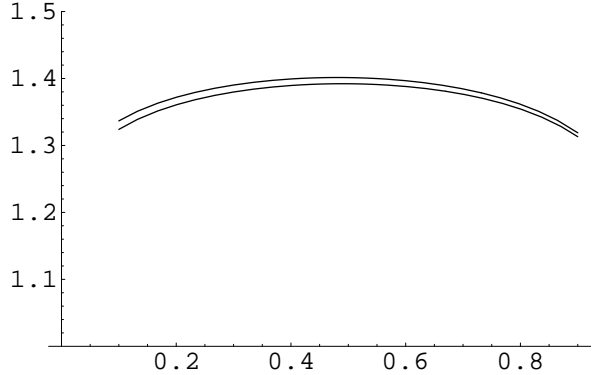}
\caption{Ratio of contribution to the total sales from 
lower $N(1-r)$ items in the ranking to that from lower $N(1-r)$ items
in the sales potential.
The upper and the lower curves correspond to $b=1.15$ and $b=1.2$,
respectively.
The horizontal and vertical axes are $r$ and $\tilde{S}(r,1)/S(r,1)$, 
respectively.}
\figa{tsg1}
\end{center}
\end{figure}
Let us return to \eqnu{totalsalestailinrankingbgt1} and
consider the role of the stochastic ranking process in inventory controls.
As an example, consider a situation where an online store
is to open a new brick-and-mortar store with $rN$ items
out of $N$ item sold at the online store.
If the manager knew the average sales rate $w_i$ of each item
$i=1,\cdots,N$ (for example, based on past records at the online store), 
he would choose the top $rN$ items
and the expected decrease in the total sales (per unit time) 
compared to the online store will be $S(r,1)$.
($w_i$ will usually be estimated based on past record of sales,
and there is a potential problem, as expressed in the Introduction,
that for items with small $w_i$, one would have small sales records,
and statistical fluctuations obscure precise determination of $w_i$
in the long tail regime. How the managers find way out in this
approach is beyond the scope of this paper.)
Now if the manager considered it quicker to select top $rN$ items 
in the \textit{ranking} at the online store, what would be the extra loss?
In this case,  the expected decrease in the total sales (per unit time) 
will be $\tilde{S}(r,1)$, so the ratio $\tilde{S}(r,1)/S(r,1)$
measures the extra loss from the use of ranking number in place of
sales rate.
\figu{tsg1} shows this ratio as a function of $r$ for $0.1\le r\le 0.9$,
calculated using \eqnu{totalsalestailinrankingbgt1}.
As a value of $b$ we adopted the values from \cite{CG,BSH}.
The ratio turned out to be insensitive to $r$ in this range
and shows 35\% to 40\% increase.
(For $r$ near $0$ and $1$, the ratio approaches $1$,
and the use of ranking data is better.
For large $b$ the ratio also approaches $1$,
and we also found that the ratio is not sensitive up to $b$ close to $1$.)
This shows an example of the use of ranking data as 
simple and effective measure of analyzing sales structure
of the long tails.

\subsubsection{Case $b<1$: The great hits economy.}
\seca{422}

Now let $b<1$ and assume $N$ is large.

As noted at the end of \secu{41}, 
when we are considering sales for $b<1$,
taking $N\to\infty$ limit results in unrealistic infinities
on average sales (sales per item), arising from divergence of great hits.
Explicitly, from \eqnu{Paretodiscrete} we have
$w_i\to\infty$ as $N\to \infty$ for each fixed $i$.
Divergence from a single item does not cause the  divergence of 
the average,
but for $b<1$, there are many such items which affect averages.

Before studying this problem,
we note that the time evolution of the ranking of a single item
which we discussed in detail in \secu{3} has no problem.
Theoretically, this reflects the fact that we assume nothing
on the distribution $\lambda$ in \prpu{yC}.
The problem of divergence of the average sales rate is 
theoretically reflected only in the fact that for
$b<1$ the Assumption (3) to \thmu{HDL} fails.
As remarked below \thmu{HDL}, this affects the distribution at $y=0$,
the top end of the rankings, but no theoretical problem occurs for $y>0$.
Intuitively speaking, if there are (fictitious) book titles which sell 
`infinitely many copies per unit time', 
they keep staying at the top end of the ranking,
and the rest of `realistic' book titles follow the evolution of
ranking as predicted by \prpu{yC}.
Also, the contribution to the total sales from the tail side 
(both $S(r,1)$ and $\tilde{S}(r,1)$ for $r>0$)
has no problem of divergence, i.e., asymptotically proportional to $N$
as in \eqnu{totalsalestailinpotentialbgt1} or
\eqnu{totalsalestailinrankingbgt1}.
In other words,
formulas not containing contributions from 
the `greatest hits' remain valid:
For $0 < r\le 1$, 
the contribution to the total sales per unit time from the
$N(1-r)$ items (out of total $N$) of low sales potentials
is as \eqnu{totalsalestailinpotentialbgt1},
$\dsp
S(r,1) \simeq \frac{Nab}{b-1} (1-r^{(b-1)/b}),
$
and that from the $N(1-r)$ items in the tail side of the ranking
is as \eqnu{totalsalestailinrankingbgt1} with \eqnu{t0yPareto},
\[
\tilde{S}(r,1)
\simeq Nab\, \Gamma(1-b,q(r))\,q(r)^{b-1};
\ \ r=1-e^{-q(r)}+ q(r)^b\, \Gamma(1-b,q(r)).
\]
\begin{figure}[hbt]
\begin{center}
\includegraphics[scale=1.0]{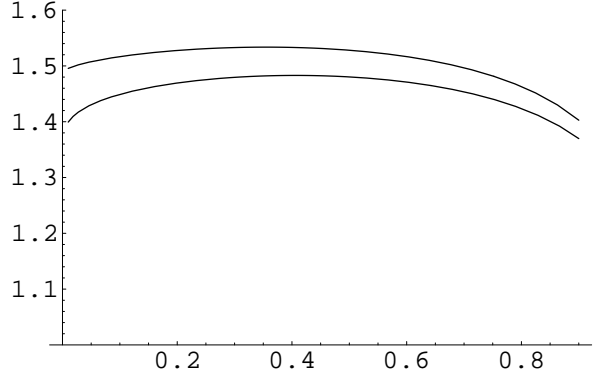}
\caption{Ratio of contribution to total sales from 
lower $N(1-r)$ items in the ranking to that from lower $N(1-r)$ items
in the sales potential.
The upper and the lower curves correspond to $b=0.6312$ and $b=0.7959$,
respectively.
The horizontal and vertical axes are $r$ and $\tilde{S}(r,1)/S(r,1)$, 
respectively.}
\figa{tsl1}
\end{center}
\end{figure}
In particular, 
we can perform a similar analysis as that concerning \figu{tsg1}
using \eqnu{totalsalestailinrankingbgt1}.
The loss in total sales per unit time caused by 
selecting top $rN$ items in the ranking
instead of selecting top $rN$ items in the sales rate
can be measured in terms of their ratio $\tilde{S}(r,1)/S(r,1)$.
\figu{tsl1} shows this ratio as a function of $r$ for $0.01\le r\le 0.9$,
calculated using \eqnu{totalsalestailinrankingbgt1}.
As a value of $b$ we adopted the values in \eqnu{low2} and \eqnu{low24}.
The ratio is below $1.6$ and insensitive to $r$ in this range.
For $r$ near $1$, the ratio approaches $1$,
and the use of ranking data is good.
(Unlike the case $b>1$ in \secu{421},
the ratio remains strictly greater than $1$ as $r\to0$.)

Returning to the problem of unrealistic infinity,
a simple modification for our approach would be to
introduce a cut off. Taking logarithms of
\eqnu{Paretodiscrete} we have
\eqnb
\eqna{Paretodiscretelog}
\log w_i=- \frac1b \log i +\frac1b \log N + \log a,\ \ i=1,2,\cdots,N.
\eqne
This formula shows that
plotting the sales rates $w_i$ against $i$  on a log--log graph,
the points will fall on a single line.
(This suggests a reason why Pareto distribution is also called
log-linear distribution and that the exponent $-1/b$ is called the 
Pareto slope parameter.)
When one assumes Pareto distributions in social and economic studies,
the argument would be in reverse direction; one probably first observes
data aligned close to a single line on a log--log graph,
and then arrive at a idealized theoretical model
\eqnu{Paretodiscretelog} or \eqnu{Paretodiscrete}.
The line actually ends in realistic situations, 
and \eqnu{Paretodiscretelog} denotes the tail end by $w_N=a$
and the head end by $w_1=a N^{1/b}$.
We let $N\to\infty$ in our formulation and as a result lost the 
head end, which causes trouble in average sales rate for $b<1$.
A simple remedy is therefore to introduce a cut-off parameter $\gamma>0$
or $n_0 = \gamma N$,
and assume a modified Pareto distribution,
\eqnb
\eqna{hitcutoffParetologlinear}
\log w_i = \log a - \frac1b \log \frac{i+n_0}{N+n_0}\,,
\ \ i=1,2,\cdots,N,
\eqne
or extend \eqnu{Paretodiscrete} as
\eqnb
\eqna{Paretodiscretewithcutoff}
w_i=a\left(\frac{N+n_0}{i+n_0}\right)^{1/b},\ \ i=1,2,3,\cdots,N.
\eqne
$\gamma=0$ or $n_0=0$ is the original Pareto distribution.
We assume Pareto distribution to be basically applicable,
so we assume $\gamma\ll 1$ ($1\ll n_0 \ll N$).

Using
\eqnu{Paretodiscretewithcutoff} in the left hand side of
\eqnu{salesr1r2}, we have
\eqnb
\eqna{totalsalestailinpotentialblt1}
\limf{N} \frac1N S(r_1,r_2)=
\frac{ab}{1-b}(1+\gamma)\left( 
(\frac{1+\gamma}{r_1+\gamma})^{(1-b)/b}
-(\frac{1+\gamma}{r_2+\gamma})^{(1-b)/b}
\right).
\eqne
If $\gamma=n_0/N=0$ we reproduce \eqnu{salesr1}.
We can safely let $r_1\to 0$ in \eqnu{totalsalestailinpotentialblt1}
 and find
\eqnb
\eqna{totalsalesheadinpotentialblt1}
S(0,r)\simeq \frac{Nab}{1-b}(1+\gamma)\left( 
(\frac{1+\gamma}{\gamma})^{(1-b)/b}
-(\frac{1+\gamma}{r+\gamma})^{(1-b)/b}
\right).
\eqne
In particular, 
\eqnb
\eqna{totalsalesblt1}
S_{tot}=S(0,1) \simeq \frac{Nab}{1-b}(1+\gamma)\left( 
(1+\frac{1}{\gamma})^{(1-b)/b}-1\right)
\simeq \frac{Nab}{1-b} \gamma^{-(1-b)/b}.
\eqne
(The left hand side is obtained by taking leading term in $\gamma\ll 1$.)
Note that we cannot let $\gamma\to 0$ for $S_{tot}$.

Other quantities can also be derived if we replace
\eqnu{Paretodiscrete} by \eqnu{Paretodiscretewithcutoff}.
Following the argument below \eqnu{Paretodiscrete},
we have, in place of \eqnu{Paretolambdadensity},
\eqnb
\eqna{Paretowithcutofflambdadensity}
\diff{\lambda}{w}(w) = \left\{ \arrb{ll} 
\dsp 0, & w> aN^{1/b}(1+\gamma^{-1})^{1/b},
\\ \dsp\frac{ba^b(1+\gamma)}{w^{b+1}}, & a< w < aN^{1/b}(1+\gamma^{-1})^{1/b},
\\ 0, & w<a. \arre \right.
\eqne
Substituting \eqnu{Paretowithcutofflambdadensity} in
\eqnu{classicstationaryexp} we have, in place of
\eqnu{classicstationaryexpPareto0},
\eqnb
\eqna{classicstationaryexpParetowithcutoff}
y_C(t)=1-b(at)^b(1+\gamma) \Gamma(-b,at)
+b(at)^b(1+\gamma) \Gamma(-b,atN^{1/b}(1+\gamma^{-1})^{1/b}).
\eqne
We note that we can take $\gamma\to 0$ limit
in \eqnu{classicstationaryexpParetowithcutoff} and
reproduce \eqnu{classicstationaryexpPareto0}.
In other words, the effect of $\gamma$ is small
for the evolution of ranking $y_C(t)$, if $\gamma$ is small.
In \secu{3} we assumed the original Pareto distribution,
and performed a fit to \eqnu{classicstationaryexpPareto}
which is equal to \eqnu{classicstationaryexpPareto0}.
That this works implies that $\gamma$ is actually small
and that  \eqnu{classicstationaryexpPareto0} is a good
approximation to \eqnu{classicstationaryexpParetowithcutoff}.
In fact, as noted at the beginning of this subsection \secu{422},
the effect of `greatest hits' on the ranking is
that they keep the top positions constantly.
The ranking data at Amazon websites are updated only once per hour,
and since there are many books which sell more than one per hour,
we never observe ranking $1$ by tracing (as we do) a book which sells
only once per months.
For such observations it is intuitively clear that taking $N\to\infty$
causes no singularities regardless of the value of $b$.

Reversing this argument, we see that since
small difference in $\gamma$ does not affect the evolution of ranking
$y_C(t)$, we cannot estimate the value of $\gamma$ from $y_C(t)$.
The dependence on $\gamma$ of the total sales $S_{tot}$ in
\eqnu{totalsalesblt1} cannot be removed,
hence for $b<1$, we cannot estimate the total sales of
the online store from the ranking data.
Our method is effective in studying the tail structures,
but is weak at great hits for $b<1$.
Standard methods, such as estimating from press reports about
top hits, should be combined, if the online store is not willing
to disclose the total sales.

Returning to \eqnu{totalsalesblt1},
we see that for $b<1$ the total sales $S_{tot}$ could be
very large (if the cut-off parameter $\gamma$ is very small) while
\eqnu{totalsalestailinpotentialbgt1} implies that
$S(r,1)$, the contribution from the tail side,
is constant in $\gamma$, hence the ratio
$S(r,1)/S_{tot}$ could be very small.
This is in contrast to the case $b>1$ discussed in \secu{421}, where 
the ratio is significantly away from $0$.
In this sense, the contribution to the sales from
the long tail would be modest in general,
and the impact of long tail business 
on economy would be also modest, if $b<1$.
Our calculations for Amazon.co.jp in \secu{3} supports $b<1$,
in spite of the Amazon group's reputation for their long tail business.
We are however aware that 
when we talk about possibility of long tail business,
there are other aspects than the contribution to the total sales or 
the direct economic impact of long tails.
For example, the phrase `the leading retail store'
is a highly effective advertisement, and being number one,
would be quoted by  mass media,
thereby drastically reduce advertisement cost.
We therefore will not be amazed if an online bookstore
takes a strategy to advertise their long tail business model,
but is hesitant about disclosing its actual sales achievement,
and makes profit largely from advance orders of `great hits' 
such as Harry Potter series.

\subsection{Conclusions.}
\seca{4conclusion}

In this paper,
we gave a mathematical framework of a new method to obtain
the distribution of sales rates of a very large number of items 
sold at an internet retail site which disclose sales rankings of
their items.
We gave explicit formulas for practical applications
and an example of a fit to the actual data obtained from Amazon.co.jp.
The method is based on new mathematical results \cite{HH071,HH072}
on a infinite particle limit of the stochastic ranking process, and is
theoretically new and quantitatively accurate.

The method is suitable especially for quantitative studies
of the long tail structure of online retails, which has been
expanding commercially
with the advance in computer networks and web technologies.
Calculation algorithm of the ranking numbers is very simple
(simplest is the best, from the theoretical side),
and will be relatively easy to implement online.
Hence our theory could serve as an efficient and inexpensive method
for disclosure policies and regulation purposes,
as well as for providing the online store business
a method of prompt analysis of long tail sales structure.
(We have heard from a book publisher that Amazon.co.jp are not
willing to open their sales results.
The publisher was amazed to know that we could estimate
Amazon's sales structure from their rankings.)
With a possible future increase in online long tail business,
the role of our theory in the business disclosure policies
may increase its  significance.

Since the result is based on mathematical results,
it is in principle applicable to general situations such as
retail stores with POS systems,
blog page view rankings, or 
the title listings of the web pages in the collected web bulletin boards.
In fact, we collected a preliminary data from 2ch.net, 
one of the largest collected web bulletin boards in Japan,
performed a fit to \eqnu{xclassicstationaryexpPareto},
and obtained a value $b=0.6145$ for the Pareto exponent, 
which is close to \eqnu{low2}. 
See  \cite{HH072} for details.
In the 2ch.net title listing page,
the titles are ordered by `the last written threads at the top' principle,
which matches the definition of the stochastic ranking process in \secu{2}.

The method would be useful for marketing
purposes as well as studies in social activities in general,
thus we consider it worthwhile to disclose the method 
for free use in practical situations.

\smallskip\par\textbf{Acknowledgements.}
We thank Prof.~K.~Takaoka for his interest in the work and
kindly giving opportunity to talk at a meeting for mathematical
finances.

The research of K.~Hattori 
is supported in part by a Grant-in-Aid for 
Scientific Research (C) 16540101 from the Ministry of Education,
 Culture, Sports, Science and Technology, and
the research of T.~Hattori 
is supported in part by a Grant-in-Aid for 
Scientific Research (B) 17340022 from the Ministry of Education,
 Culture, Sports, Science and Technology.

\end{document}